\documentclass[psfig]{aa}                                                    
\input psfig.tex

\begin{document}                                                                
\def\et{et al.}                                                                 
\def\egs{erg s$^{-1}$}                                                          
\def\egsc{erg s$^{-1}$ cm$^{-2}$}                                               
\def\msu{M$_{\odot}$\ }                                                         
\def\kms{km s$^{-1}$ }                                                          
\def\kmsM{km s$^{-1}$ Mpc$^{-1}$ }                                              
   \thesaurus{06         
              (03.11.1)}  

   \title{XMM-Newton observations of M87 and Its X-Ray Halo}         
                                                                                
   \author{H. B\"ohringer \inst{1}, E. Belsole \inst{2}, J. Kennea  \inst{3},
           K. Matsushita \inst{1}, S. Molendi  \inst{4},  
           D. M. Worrall \inst{5}, R.F. Mushotzky \inst{6}, M. Ehle, \inst{7}
           M. Guainazzi \inst{7}, I. Sakelliou \inst{8}, 
           G. Stewart  \inst{9}, W.T. Vestrand \inst{10}, S. Dos Santos \inst{9} 
   }

   \offprints{H. B\"ohringer}                                                   
                                                                                
   \institute{$^1$ Max-Planck-Institut f\"ur Extraterrestrische Physik,         
                   D-85748 Garching, Germany\\                                  
              $^2$ CEA Saclay, Service d'Astrophysique, 91191 Gif sur Yvette, France\\
              $^3$ Department of Physics, University of California, Santa Barbara, CA 93106, USA\\
              $^4$ Istituto di Fisica Cosmica, Via Bassini 15, I-20133 Milano,
                   Italy \\
              $^5$ Department of Physics, University of Bristol, Tyndall Avenue,
                      Bristol BS8 1TL, UK\\          
              $^6$ Laboratory for High Energy Astrophysics, Code 660, 
                   NASA/Goddard Space Flight Center, Greenbelt, MD 20771, USA\\
              $^7$ XMM-Newton SOC, VILSPA -ESA, PO Box/Apartado 50727, E-28080 Madrid, Spain \\
              $^8$ Mullard Space Science Laboratory, University College London, 
                   Holmbury St Mary, Dorking, Surrey RH5 6NT\\
              $^9$ Department of Physics and Astronomy, The University of
                  Leicester, Leicester LE1 7RH\\
              $^{10}$ NIS-2, MS D436, Los Alamos National Laboratory, USA
                    } 
                                                                                
   \date{Received ..... ; accepted .....}                                       
                                                                                
   \maketitle                                                                   
                                                                                
   \markboth {XMM-Newton Observations of M87}{}                                  
                                                                                
\begin{abstract}                                                                
We report performance verification observations of the giant 
elliptical galaxy 
M87 in the Virgo Cluster with the MOS, pn, and optical monitor 
instruments on board of XMM-Newton.
With the energy sensitive imaging instruments MOS and pn we obtain 
the first spatially constrained X-ray spectra of the nucleus 
and the jet of the galaxy. The good photon statistics of the pn 
and MOS allow a detailed analysis of the radial temperature 
and abundance distribution of 6 elements. The data provide no 
indication of a multi-temperature structure for radii $\ge $2$'$.
An apparent sharp metal abundance drop deduced for the
regions inside this radius is probably due to 
resonant line scattering.


   \end{abstract}                                                               
                                                                                
%
                                                                                
\section{Introduction}                                                          
                                                                                
The giant elliptical galaxy M87 as the central dominant galaxy of the
northern part of the Virgo cluster (e.g. Binggeli et al. 1987)
displays a large variety of very interesting astrophysical phenomena
(see e.g. R\"oser \& Meisenheimer 1999). 
Because M87 is so bright and close, 17 - 20 Mpc (Freedman et al. 1994,
Tammann \& Federspiel 1997),
these astrophysical processes can be studied in M87 in more detail than
in any other comparable object.

M87 (Virgo A, 3C 274) is one of the first known
extragalactic radio sources and its halo is the 
first extragalactic X-ray source
to have been identified (Bolton \& Stanley 1948, Byram et al. 1966). The
highly peaked surface brightness profile of the extended 
X-ray emission subsequently observed with EINSTEIN 
implied an X-ray gas cooling time considerably smaller than the Hubble time, 
leading to the suggestion that M87 harbours a significant ``cooling
flow'' with a mass condensation rate of $\sim 10$ M$_{\odot}$ yr$^{-1}$
(Stewart et al. 1984, Fabian et al. 1984). 
EINSTEIN Crystal Spectrometer and Solid State Spectrometer
observations of M87 provided 
X-ray line data which lent further support to the existence of
a multi-temperature plasma in the core of the M87 halo, as
expected for a cooling flow model (Canizares et al. 1979, 1982,
Mushotzky \& Szymkowiak 1988).

M87 also houses an active nucleus with a one-sided jet observed
at radio (e.g. Owen 1989), optical (e.g. Sparks et al. 1996), 
and X-ray wavelengths (Schreier et al. 1982, Neumann et al. 1997,
Harris et al. 1997, 1999).
The power output of the jet and the unseen counterjet obviously
feeds a complex system of inner ($r \le 25$ kpc)
and outer radio lobes ($r \le 40$ kpc, e.g. Owen et al. 2000)
which partly interact
with the thermal gas giving rise to distinct
features in the X-ray surface brightness distribution 
(e.g. Belsole et al. and references therein). 
The ability to obtain sensitive spatially resolved X-ray spectroscopy 
with the XMM-Newton observatory  
offers the opportunity to gain new insights into these
phenomena. 
            
This paper reports on the analysis of the XMM-Newton (Jansen et al. 2000) 
performance verification phase observation of M87 
including results from the MOS (Turner et al. 2000), pn
(Str\"uder et al. 2000) and Optical Monitor (Mason et al. 2000) instruments.
An analysis of the
morphology and spectroscopy of the X-ray surface brightness
enhancements related to the radio lobes of M87 is presented in
an accompanying paper by Belsole et al. (2000), and the results
from high resolution spectroscopy with the RGS instrument
are discussed in a forthcoming paper by Sakelliou et al. (2000).

                                                                                
\section{XMM observations}                                                    
                                                                                
M87 was observed with XMM-Newton 
on June 19th, 2000 for 75.6 ksec. The effective
exposure of the pn detector with the thin filter is 25.9 ksec,
with no signes of enhanced background events.
For the data reduction the 
SAS software was used as available in August 2000. 
Fig. 1 shows a full frame image of the pn-detector observation
in the energy range 0.5 to 2 keV. 
The effects of out-of-time events registered
during the reading of the camera chips can be seen as a thin luminous
band in the lower chip, fourth from the left. The X-ray halo has an
almost spherically symmetric appearance, 
except for two localized
surface brightness enhancements to the SW and E of the M87 nucleus. 
These features coincide with the inner radio lobes. 
In total more than 12 Million photon events were registered 
by this detector.
For the MOS1 and MOS2 detectors an effective observing time of
32 ksec is obtained after excluding the first 7 ksec
due to enhanced background in the 10 - 12 keV band.

\begin{figure}                                                                  
\psfig{figure=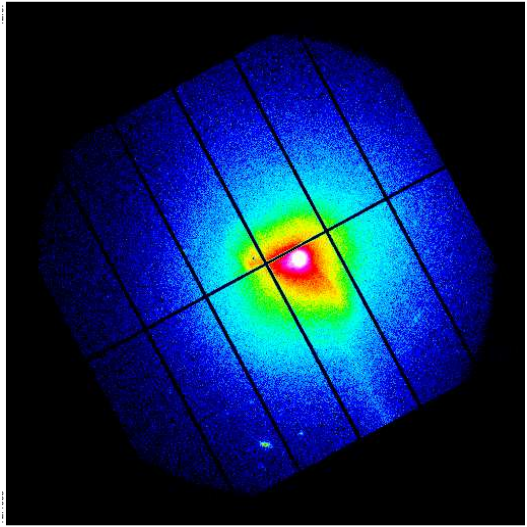,height=7cm}                                        
\caption{XMM-Newton pn image of M87 in the energy range 0.5 to 2 keV.
The image size is about $26'$ by $27'$, north is up and east is left.}                                                                               
\end{figure}

\section{The Nucleus and Jet of M87}
                  
M87's nucleus and famous jet as seen with the Optical
Monitor camera with the UWV2 filter is shown in Fig. 2. A cross section
of the deconvolved surface brightness distribution along the nucleus-jet
direction is shown in Fig. 3. Several of the bright knots in the 
jet are easily visible. Interestingly, knot A in the jet
is brighter than the nucleus (see Sparks et al. 1996). The observed count rate
for the nucleus is 0.290 cts s$^{-1}$ corresponding to a flux
of $0.190 \pm 0.004$ mJy at 2120 $\AA$.

\begin{figure}
\psfig{figure=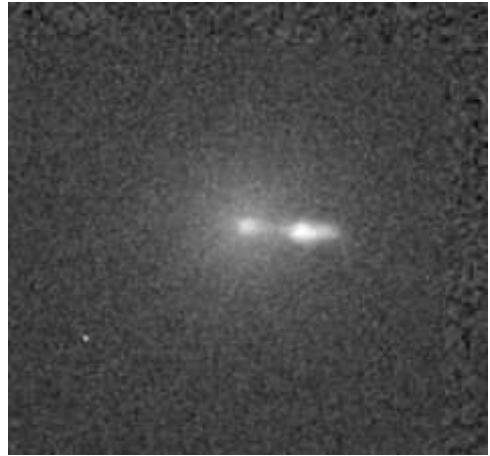,height=6cm}
\caption{UV image of the nucleus and the jet of M87 observed with
the optical monitor and the UVW2 filter ($\sim 180 - 230$ nm).
North is not up in this image.
}
\end{figure}

\begin{figure}
\psfig{figure=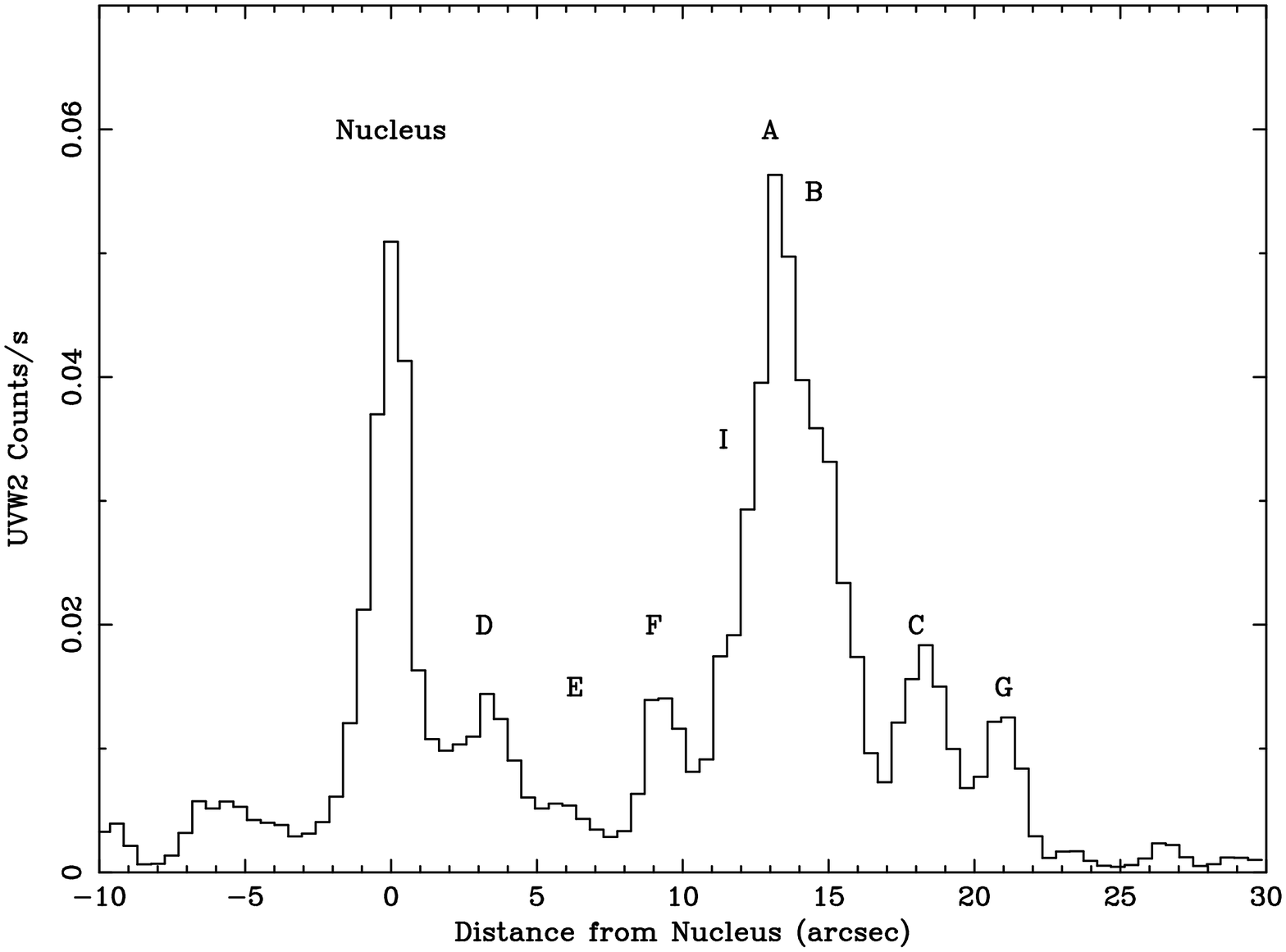,height=6cm}
\caption{Cross section through the deconvolved surface brightness of the
UV image of the nuclear region in the line connecting the nucleus
and the jet.}
\end{figure}

Fig. 4 shows the MOS2 X-ray image of the nuclear region. The
nucleus and the jet appear as two point sources
separated by about 11.5 arcsec  (with a measurement uncertainty
of about 0.5 arcsec).
This is close to, but not exactly the same as, the nucleus to 
knot A separation of 13$''$ seen in Fig. 3. This difference was
already noted by Neumann et al. (1997) who quote a core-jet separation of
11.9  and 12.7 arcsec for the X-ray (ROSAT HRI) and radio
image, respectively (1$''$ corresponds to 82 pc at the adopted 
distance to M87 of 17 Mpc). The (2 - 10 keV) flux for the jet 
and the nucleus is about $5.5 \times 10^{-13}$ erg s$^{-1}$ cm$^{-2}$ and 
$1.5 \times 10^{-12}$ erg s$^{-1}$ cm$^{-2}$, respectively.

\begin{figure}
\psfig{figure=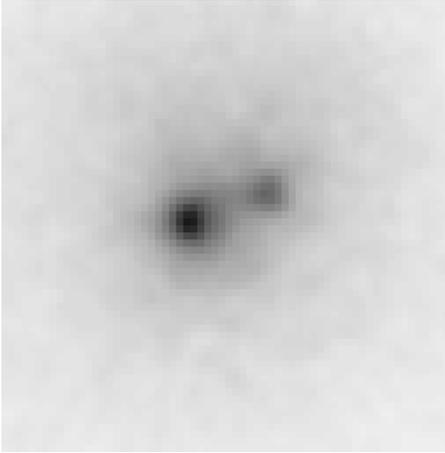,height=6cm}
\caption{MOS1 X-ray image image of the nucleus and the jet of M87
in the energy range 0.3 to 5 keV.
The side length of the image is 63 arcsec. The image was smoothed 
with a Gaussian filter with a $\sigma $ of 1 arcsec}
\end{figure}

XMM-Newton allows us for the first time to obtain separate X-ray 
spectra for the nucleus and X-ray knot in the jet. For the analysis
of the pn data
we have extracted spectra from circular regions around each
with a radius of 5 arcsec and taken two neighbouring regions of the same size
for the background subtraction, one 20.8 arcsec in NNW and one 17.7 arcsec
in SWW direction. The resulting spectra obtained using the first
background region are shown in 
Figs. 5 and 6. The spectra can be well fitted (in the 0.2 to 6 keV band)
by power law spectra with
a slope of $2.2 (\pm 0.2)$ for the nucleus and about $2.5 (\pm 0.4)$
for the jet.  
There is no significant signature of thermal emission
(which can qualitatively
be noted in Figs. 5 and 6 by an absence of a blended peak of iron lines
around 1 keV in the residual spectra).
 
Similar analysis was performed with the combined MOS data using
an extraction radius of 4 arcsec for the nucleus and 3.5 arcsec for
the jet and a background region at 22 arcsec
distance from the nucleus. The results for the power low slope are 
$2.25 {+0.04 \brack -0.06}$ for the nucleus and $2.45 
{+0.11 \brack -0.08}$ for the 
jet (with acceptable fits with $\chi ^2$ of 170/178 d.o.f. and 101/110 d.o.f).
A pure thermal origin of the jet emission can be ruled out with high
confidence.

\begin{figure}
\psfig{figure=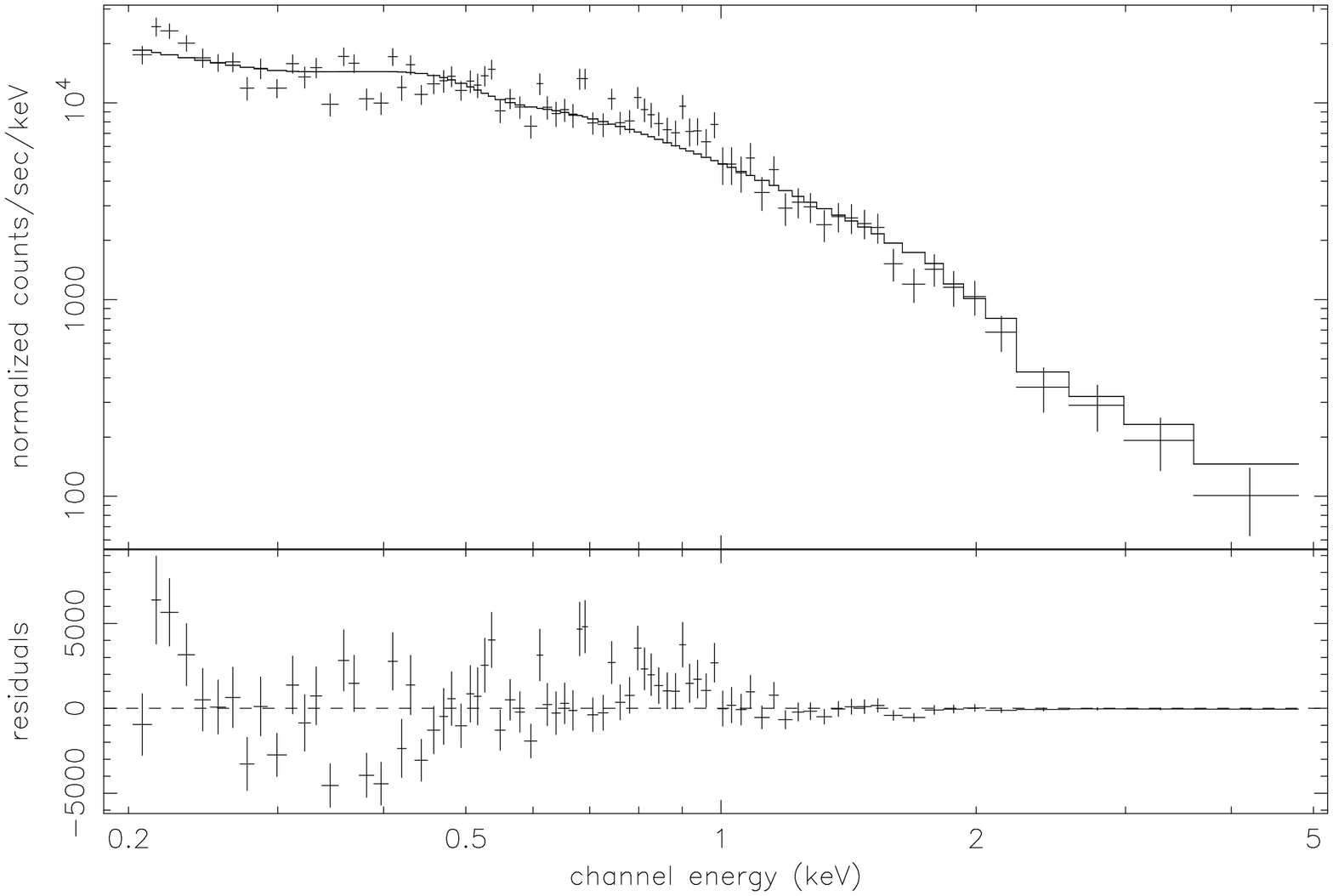,height=5.5cm}
\caption{XMM pn spectrum of the nucleus of M87. The solid line shows
a fit of an absorbed (galactic value) power law spectrum 
with a slope of 2.2.} 
\end{figure}

\begin{figure}
\psfig{figure=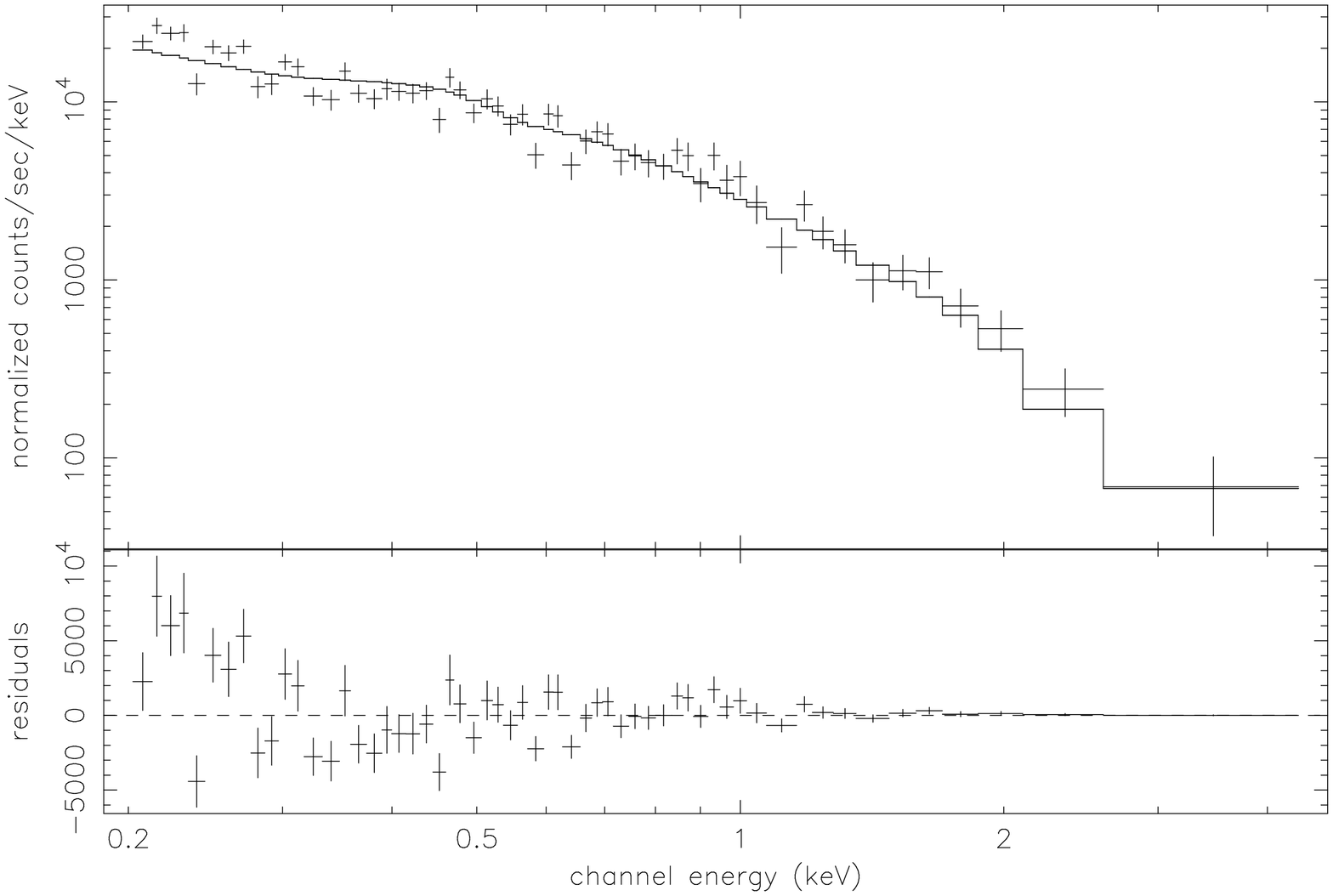,height=5.5cm}
\caption{XMM pn spectrum of the X-ray knot in the jet of M87.
The solid line shows
a fit of an absorbed (galactic value) power law spectrum 
with a slope of 2.5.}
\end{figure}

Fig. 7 shows the radio to optical spectral energy distribution of 
the jet region encompasing knots A, B, and C 
(Biretta et al. 1991, Meisenheimer et al. 1996, Owen et al. 1989,
Perola et al. 1980, Stocke et al. 1981, Tansley, 2000) together
with the XMM-Newton results.

\begin{figure}
\psfig{figure=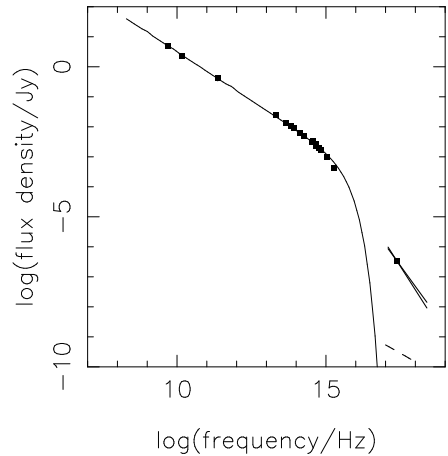,height=6cm}
\caption{Spectral energy distribution of knot A, B, and C of the jet in M87
compared to a model of synchrotron emission
(solid line) and self-Compton emission (broken line).  The present observational
result is marked by the data point and the two short lines indicating the lower
and upper bound on the power law slope, respectively.}
\end{figure}

\section{Temperature Structure of the Intracluster Medium}

XMM-Newton allows us for the first time to
analyse X-ray spectra in many regions of the M87 X-ray halo with
good photon statistics. Since the X-ray halo appears almost
azimuthally symmetric, we study spectra in a set of concentric
rings around the nucleus of M87. 
 We exclude the regions of excess emission at the location of the
radio lobes  (a region of $4'$ by $4'$ centered
on a point $2'$ south and $2'$ west and 
a region of $4.5'$ by $4'$  $2'$ east and
$0.5'$ north of the nucleus, respectively) and
the regions where the fraction of out of time events is large.  
We use the preliminary response matrix provided in July 2000  
and an arf file calculated from vignetting data provided by
B. Aschenbach (private communication). For the background,
70 ksec of the Lockman hole observation in orbit 70, 71, and 73
cleaned of bright sources and background flares is used.
A second analysis was performed with the cleaned MOS1 data.
Vignetting corrected spectra were extracted from the event
list. For the background a combination of several deep exposures
collected by D. Lumb (private communication) where sources have 
been removed is used.

\begin{figure}
\psfig{figure=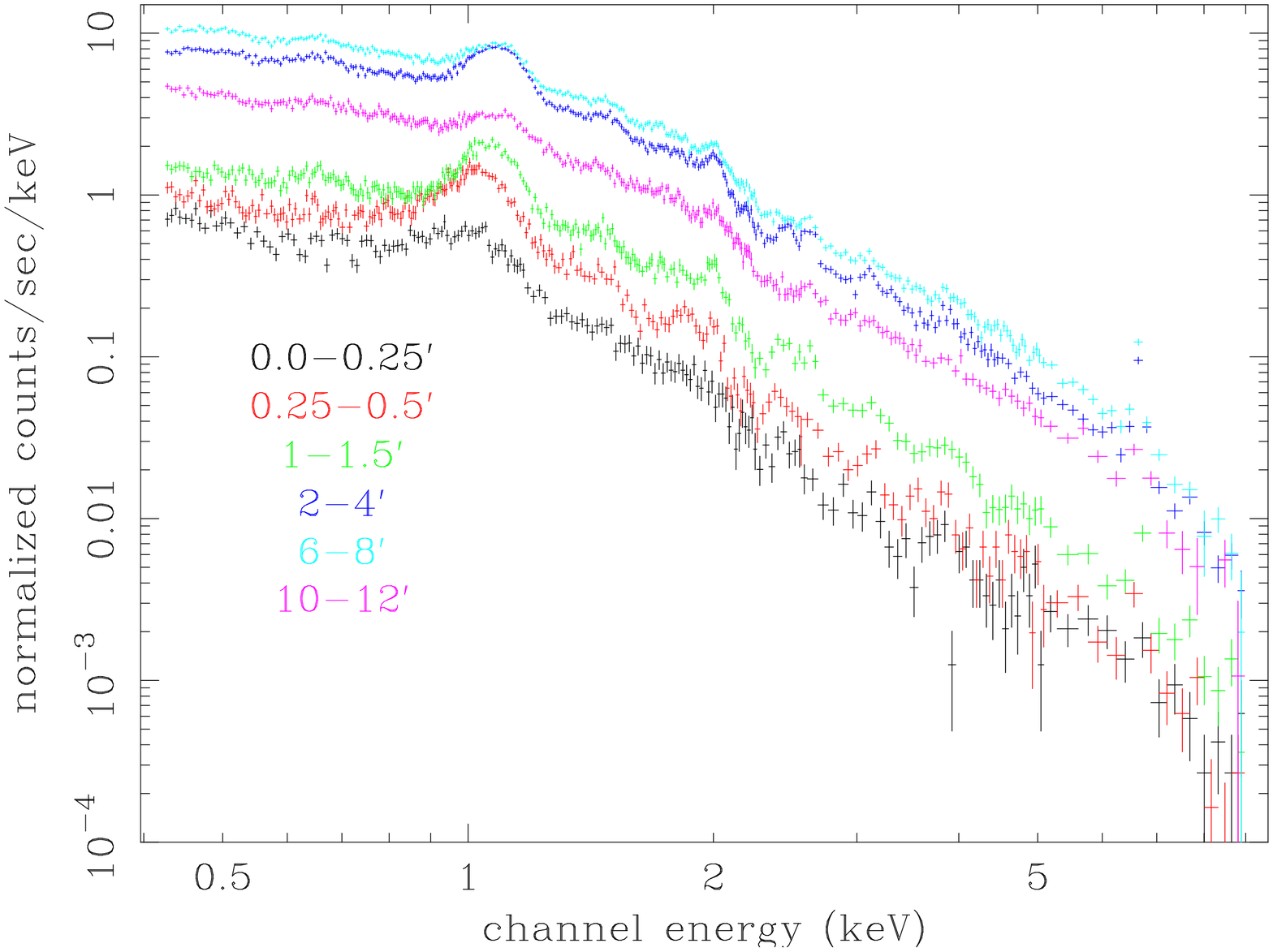,height=6cm}
\caption{pn spectra of the X-ray emission in concentric rings
around M87. Clearly seen is the variation of the emission line
strength as a function of radius, indicating abbundance variations.}
\end{figure}

Resulting pn spectra are shown in Fig.
8. The fitting of model spectra was performed in XSPEC using
the vmekal model. For all rings outside a radius of 1 arcmin 
two-temperature models provided no significant improvement over
the fits of single-temperature models in the analysis of results of
both detectors. For the inner circle a power law component was added
to the fit. 
Figs. 9 to 12 show the results of single temperature
model fits for the pn and MOS1 data.
Results of the different detectors agree well,
except for some difference in the determination of the oxygen
abundance (Fig. 12). 

\begin{figure}                                                                  
\psfig{figure=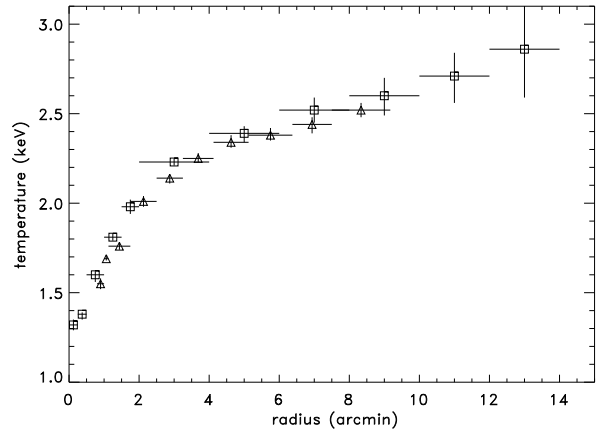,height=6cm}                                        
\caption{Temperature profile determined from the MOS (triangles) 
and pn spectra (squares) in concentric rings around the M87 nucleus 
excluding the regions of the X-ray lobes.
}                                                                               
\end{figure}

The temperature profile shown in Fig. 9 provides a very tight constraint
on the temperature distribution in the M87 halo and it is in perfect
agreement with the ASCA results from Matsumoto et al. (1996) 
and the BeppoSAX results of Guainazzi and Molendi (2000). 
The temperatures are somewhat higher than determined
with the ROSAT PSPC observation (Nulsen \& B\"ohringer 1995)
but the trend with radius is similiar.

\section{Element abundances}

A comparison of the element abundances with the ASCA results of 
Matsumoto et al (1996) shows reasonable agreement for sulfur and the higher
oxygen results obtained with the pn detector, while the ASCA 
results for silicon are on average about 30\% and the iron
abundances about 10-20\% lower.  The equivalent width for the Si
lines are in good agreement, however, and therefore the differences
in the abundances may solely be due to the different plasma codes 
used (rs and vmekal). 

In summary the profiles can be characterized by decreasing abundances
with radius outside a radius of 1.5 arcmin, except for oxygen which 
is almost constant. Notable is the dramatic
abundance decrease towards the center inside a radius of 1 - 1.5 arcmin.
We have further tested the reality of the decrease by
considering two-temperature spectral models for the innermost three
rings. For the best fitting models the upper temperature is only 
slightly higher than the temperature of the one-component fits
and the lower temperature is roughly half this value. As shown in
Fig. 10 for the examples of iron and silicon, the abundance decrease 
towards the center is smaller in the multi-component models
than in the one-temperature fits, but it is still quite
striking.

\begin{figure}
\psfig{figure=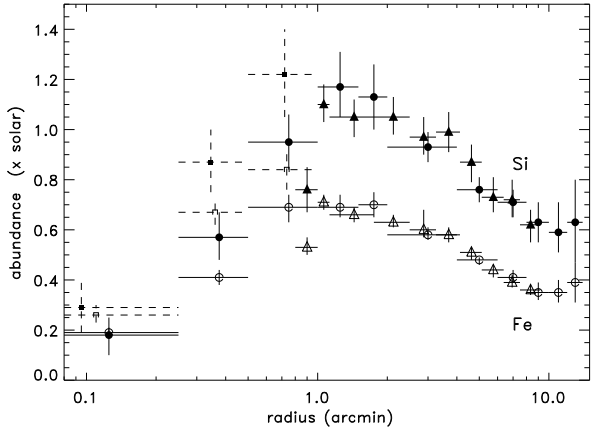,height=6cm}
\caption{Abundance profiles of Si (solid symbols) and Fe (open symbols)
determined from the MOS1 (triangles)
and pn spectra (circles) in concentric rings around the M87 nucleus
excluding the regions of the X-ray lobes. The data points with dashed
error bars show the results of two-temperature fits to pn-spectra
for Si (solid squares) and Fe (open squares). 
}
\end{figure}

\begin{figure}
\psfig{figure=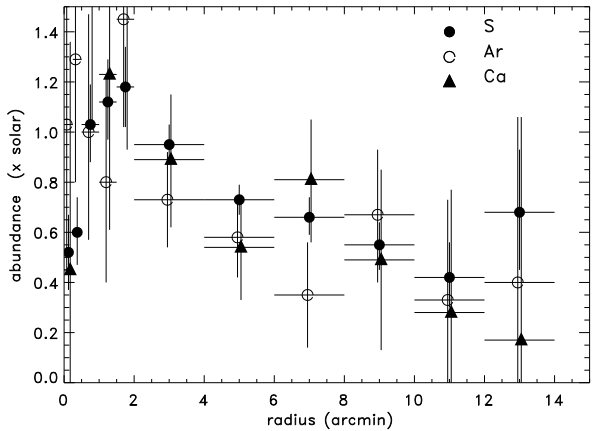,height=6cm}
\caption{Abundance profiles of S, Ar, Ca determined from 
pn spectra in concentric rings around the M87 nucleus
excluding the regions of the X-ray lobes.
}
\end{figure}

\begin{figure}
\psfig{figure=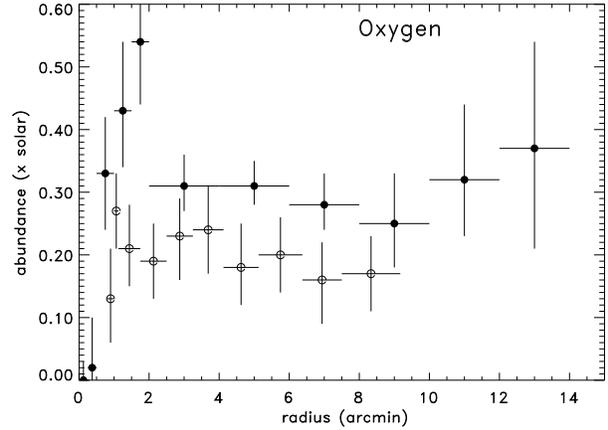,height=6cm}
\caption{Abundance profile of Oxygen determined from
MOS1 (open circles) and pn spectra (solid circles)
in concentric rings around the M87 nucleus
excluding the regions of the X-ray lobes.
}
\end{figure}

\section{Resonant line scattering}

As discussed previously (e.g. Gil'fanov et al. 1987, Tawara et al. 1997,
Shigeyama 1998) resonant scattering of line emission should be
important in dense cores of galaxy clusters.
This effect might well be the cause of the apparant abundance decrease
towards the cluster center. To illustrate the effect we show in Fig. 13
the results of simple model calculations of the optical depth of 
some important resonance lines for an isothermal plasma halo  
with a temperature of 1.3 keV and abundances
as found for the radial range 1 to 1.5 arcmin.
The main difference to previous
model calculations (e.g. Shigeyama 1998) is the very small
core radius (20$''$, 1.97 kpc) used here as indicated by
the ROSAT HRI observations (e.g. B\"ohringer 1999).
We note that all the Ly$\alpha $
lines of the H-like ions, the resonant lines of the He-like ions
and, many important Fe L-shell lines become moderately optically 
thick at the center. Since the optical thickness
is moderate in the center
and the density profile is quite steep, 
most of the last scattering must occur at a few core radii
($r= 1'$ to 2$'$). In the most 
extreme case the effect will result in a flat
surface brightness profile of the lines inside this radius.
Since the abundances reflect the ratio
of the line emission to the continuum emission and 
since the continuum surface brightness increases by a factor
of about 8 - 10 from a radius of 1 to 1.5 arcmin to the center,
we expect at most a drop in the apparent abundances by the same factor.
Therefore it seems that the results are very plausibly explained
by resonant line scattering. Further modeling and a careful investigation
of the X-ray spectra is necessary to provide a more detailed proof.

\begin{figure}
\psfig{figure=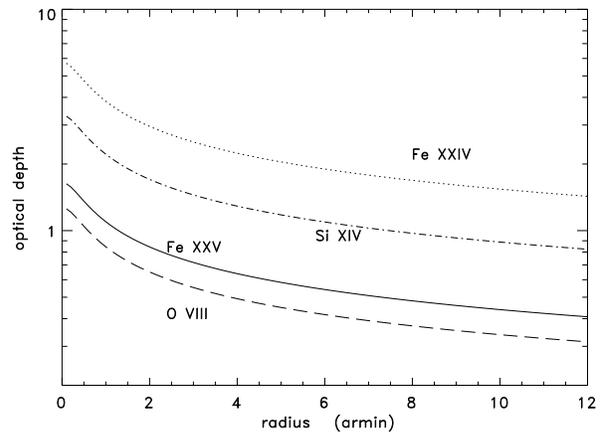,height=6cm}
\caption{Optical depth for resonant line scattering of some prominent
emission lines of Fe, Si, and O observed in the M87 spectrum. For the
plasma distribution we used a beta model with the parameters: core radius
$= 1.97$ kpc, $\beta = 0.47$, central electron density $ = 0.35$ cm$^{-3}$, 
and an isothermal temperature of 1.3 keV. The lines are labled by
the relevant ions and refer to Fe XXV (6.69 keV), Fe XXIV (1.16 keV),
Si XIV (2.003 keV, Si Ly$\alpha $), O VIII (0.652 keV, O Ly$\alpha $). 
}
\end{figure}

\section{Evidence for multi-temperature structure}

The spectral fitting results for concentric rings described above  
did not show any improvement by including more
than one temperature component, except for the rings inside about 1 - 1.5.
This is in contrast to current cooling flow
models for which we would expect to see a multi-phase medium 
out to a larger radius and a larger temperature range for the inner
region.
To further test the consistency with cooling flow models we performed 
combined model fits of a single temperature and cooling flow model
(vmekal and mkcflow in XSPEC) for a set of circular regions around
the M87 center. Fig. 14 shows the results for the two cooling flow
parameters, mass deposition rate and internal absorption column density,
as a function of the outer radius of the circular region. We
expect to observe a decreasing absorption column with radius for
the cooling flow model, we observe an increase, however. 
If we set the internal absorption
to zero in the fit,  the mass deposition rate is reduced to a value
below 1 M$_{\odot}$ per year. We interpret these results
as inconsistent with the conventional cooling flow model, and the
increasing mass deposition rate observed in Fig. 14 is forced onto
the fit, because spatially separated temperatures in different rings
are observed simultaneously in the spectra extracted from circular
regions.  

\begin{figure}
\psfig{figure=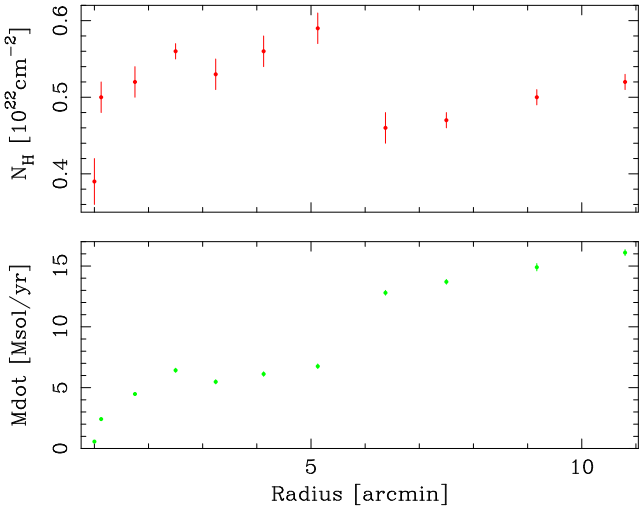,height=7cm}
\caption{Results of a cooling flow analysis in concentric circles
around M87 (using MOS1). Shown are the parameters for the warm absorption 
hydrogen column density in the cooling flow and the infered
mass flow rate (for details see text). }
\end{figure}

\section{Discussion and Conclusions}

In Fig. 7 the spectral energy distribution of the bright knots in the 
jet are compared with a model for synchrotron radiation 
(from an electron spectrum of slope 2.3, with $\gamma_{\rm min} = 100$ 
and $\gamma_{\rm max} = 2 \times
10^6$ in the equipartition magnetic field of 265 $\mu $Gauss). 
Also shown is the contribution of self-Comptonization for an
equipartition field. Both components fall short of explaining the 
flux observed with XMM-Newton. The observations could be matched by
synchrotron emission from an electron spectrum which extends to higher
energies above a break at $\gamma \sim 8 \times 10^5$, but the 
required steepening of the 
electron spectral slope of $\sim 1.5$ is not naturally predicted by
simple models. The high electron energies and consequently short lifetimes
imply that the electrons are accelerated in-situ at the X-ray knot
(see also Heinz \& Begelman 1997). 
Lifetime effects could then explain why the optical and X-ray knots 
appear in slightly different positions.

The good agreement of the temperature profiles between the different
XMM-Newton instruments as well as the ASCA and BeppoSAX instruments
is a sign that the overall calibration of the detector response is
already very good. 

One of the surprising results is the non-detection of good evidence
for a multi-temperature cooling flow structure for which M87 was 
expected to be a good target. We also do not find the signature of 
oxygen absorption claimed to be seen by Buote (2000) in ROSAT PSPC data. 

A striking result is the sharp drop of all deduced metal 
abundances at radii smaller than about 1 arcmin. This effect 
can be observed directly in Fig. 8. We have shown that 
resonant line scattering is important and could
be responsible for this effect. This effect
has also to be taken into account in the interpretation of 
XMM-RGS spectra of nearby clusters, since some of the 
relevant emission lines may suffer from scattering out-off the 
field-of-view of the spectrometer. 

The decrease of the abundance profiles in the outer regions 
show a significant trend with element mass. The decrease
from $r = 1.5'$ to $r = 10'$ is about a factor of
2 or larger for Fe, S, Ar, less for Si (a factor of $\sim 1.7$),
and the profile is almost constant for O. This can be explained 
by a contribution to a homogeneous distribution by SN type II,
the main source for O, and a SN type Ia yield of primarily heavy
elements which is more concentrated in the center of the M87 halo.

\begin{acknowledgements}                                                        
We thank the XMM software team for providing the Software Analysis System (SAS)
for the XMM-Newton data reduction. H.B. thanks D. Grupe, F. Haberl
for help with the SAS data analysis and
E.M. Churazov for helpful comments. 
We are grateful to Martin Hardcastle 
for model-fitting software used to produce the model shown in Fig. 7.
The paper is based on observations obtained with XMM-Newton, an ESA
science mission with instruments and contributions directly funded by
ESA Member States and the USA (NASA). The XMM-Newton project is supported
by the Bundesministerium f\"ur Bildung und Forschung, Deutsches Zentrum 
f\"ur Luft und Raumfahrt (BMBF/DLR), the Max-Planck Society and the
Haidenhain-Stiftung. 

\end{acknowledgements}

\end{document}